%% file: main.tex
\documentclass[finalrunningheads,a4paper]{llncs}

\input{styles.tex}

\input{macro.tex}

\newcommand{\solvent}{Solvent\xspace}
\newcommand{\solventgithub}{https://github.com/AngeloFerrando/Solvent}
\newcommand{\experimentsgithub}{https://github.com/AngeloFerrando/Solvent/tree/main/results}
\newcommand{\vertasks}{107\xspace}

\title{\solvent: liquidity verification of smart contracts}

\newtoggle{arxiv}
\toggletrue{arxiv}

\begin{document}

\pagestyle{plain}

\renewcommand{\thelstlisting}{\arabic{lstlisting}}

\newcommand{\emailaddress}{enrico.lipparini@edu.unige.it}

\author{Massimo Bartoletti\orcidlink{0000-0003-3796-9774
}\inst{1}, Angelo Ferrando\orcidlink{0000-0002-8711-4670}\inst{2}, Enrico Lipparini\orcidlink{0009-0009-0428-4403}\href{mailto:\emailaddress}{$^{\textrm{(\Letter)}}$}\inst{1, 3}, Vadim Malvone\orcidlink{0000-0001-6138-4229}\inst{4}}

\institute{
Universit\`a degli Studi di Cagliari, Cagliari, Italy
\and 
Universit\`a degli Studi di Modena e Reggio Emilia, Modena, Italy
\and 
Universit\`a degli Studi di Genova, Genova, Italy
\and 
T\'el\'ecom Paris, Palaiseau, France}

\maketitle

\input{abstract.tex}

\input{intro.tex}
\input{liqVsSafLive.tex}
\input{tool.tex}
\input{evaluation.tex}
\input{related.tex}
\input{conclusions.tex}

\input{ack.tex}

\bibliographystyle{splncs04}
\bibliography{main}

\iftoggle{arxiv}{
\newpage
\appendix
\input{app-intro.tex}
\input{solidity.tex}
\input{app-related.tex}
\input{lottery.tex}
}{}

\end{document}

%% file: styles.tex
\usepackage[T1]{fontenc}

\usepackage{color}
\usepackage[usenames,dvipsnames]{xcolor}

\usepackage[framemethod=tikz]{mdframed}

\usepackage{multicol}
\usepackage{booktabs}
\usepackage{siunitx}
\usepackage{textcomp}

\usepackage[most]{tcolorbox}

\tcbset{myoneside/.style={
colback=LightGrey,
colframe=gray,
top=0pt,
left=2pt,
right=2pt,
bottom=-5pt,
boxsep=1pt,
halign=flush center,
nobeforeafter,
before skip=-5pt,
after skip=0pt,
boxrule=1pt
}}

\tcbset{mytwosides/.style={
colback=LightGrey,
colframe=gray,
top=0pt,
left=2pt,
right=2pt,
bottom=-5pt,
boxsep=1pt,
sidebyside,
sidebyside align=top,
sidebyside gap=8pt,
halign=flush center,
nobeforeafter,
before skip=-5pt,
after skip=0pt,
boxrule=1pt
}}

\usepackage{pst-func}
\usepackage{amssymb}
\usepackage{pifont}
\usepackage{graphicx}
\usepackage{svg}

\usepackage{environ}

\usepackage{tikz}
\usepackage{forest}
\usetikzlibrary{shadows}
\usetikzlibrary{matrix,chains,positioning,decorations.pathreplacing}
\usetikzlibrary{patterns,calc,fit,arrows}
\usetikzlibrary{shapes.geometric}
\usetikzlibrary{positioning,arrows.meta,backgrounds}

\usepackage{mathtools}
\usepackage{array,multirow}
\usepackage{algorithm}

\usepackage{mdframed}
\usepackage{arydshln}
\usepackage{etoolbox}
\usepackage{xspace} 
\usepackage{dirtytalk} 
 
\usepackage{makecell}
\usepackage{stackengine}
\setstackEOL{\cr} 

\usepackage[inline,shortlabels]{enumitem} 
\newlist{inlinelist}{enumerate*}{1}
\setlist*[inlinelist,1]{%
  label=(\roman*),
}

\usepackage{nicefrac}

\usepackage{xifthen}  
\usepackage{ifthen}

\usepackage[hidelinks]{hyperref}
\usepackage{cleveref}
\usepackage{bigints}
\hypersetup{final} 

\usepackage[final,nomargin,inline,index]{fixme} 
\fxusetheme{color}

\FXRegisterAuthor{bart}{anbart}{\color{magenta} {\underline{bart}}}

\usepackage{hhline}
\usepackage{array,multirow}
\usepackage[disable]{todonotes}

\hypersetup{
  breaklinks   = true,
  colorlinks   = true, 
  urlcolor     = blue, 
  linkcolor    = blue, 
  citecolor    = red   
}
\hypersetup{final} 

\usepackage{caption}
\DeclareCaptionType{listing}[Listing][List of Listings] 
\crefname{listing}{Listing}{Listings}  

\usepackage{xurl} 

\usepackage{listings}

\usepackage[misc]{ifsym}
\usepackage{orcidlink}

\definecolor{listingBG}{HTML}{FFFFCB}%
\definecolor{listingFrame}{HTML}{BBBB98}%
\definecolor{listingLineno}{rgb}{0.5,0.5,1.0}%

\definecolor{LightGrey}{rgb}{0.975,0.975,0.975}

\lstdefinelanguage{solidity}{
	commentstyle=\color{Gray},
	morecomment=[l]{//},
	morecomment=[s]{/*}{*/},
	classoffset=0,
        escapechar=\$,
	morekeywords={property, anonymous, assembly, balance, break, call, callcode, case, catch, class, constant, continue, constructor, contract, debugger, default, delegatecall, delete, do, else, emit, event, experimental, export, external, false, finally, for, function, gas, if, implements, import, in, indexed, instanceof, interface, internal, is, length, library, log0, log1, log2, log3, log4, memory, modifier, new, payable, immutable, pragma, private, protected, public, pure, push, return, returns, revert, rule, selfdestruct, send, solidity, storage, struct, suicide, super, switch, then, this, throw, transfer, true, try, typeof, using, value, view, while, with, addmod, ecrecover, keccak256, mulmod, ripemd160, sha256, sha3},
	keywordstyle=\color{NavyBlue}\bfseries,
	classoffset=1,
	morekeywords={address, bool, byte, bytes, bytes1, bytes2, bytes3, bytes4, bytes5, bytes6, bytes7, bytes8, bytes9, bytes10, bytes11, bytes12, bytes13, bytes14, bytes15, bytes16, bytes17, bytes18, bytes19, bytes20, bytes21, bytes22, bytes23, bytes24, bytes25, bytes26, bytes27, bytes28, bytes29, bytes30, bytes31, bytes32, enum, int, int8, int16, int24, int32, int40, int48, int56, int64, int72, int80, int88, int96, int104, int112, int120, int128, int136, int144, int152, int160, int168, int176, int184, int192, int200, int208, int216, int224, int232, int240, int248, int256, mapping, string, uint, uint8, uint16, uint24, uint32, uint40, uint48, uint56, uint64, uint72, uint80, uint88, uint96, uint104, uint112, uint120, uint128, uint136, uint144, uint152, uint160, uint168, uint176, uint184, uint192, uint200, uint208, uint216, uint224, uint232, uint240, uint248, uint256, var, void, ether, finney, szabo, wei, days, hours, minutes, seconds, weeks, years},
	keywordstyle=\color{blue},
	classoffset=2,
	morekeywords={block, blockhash, coinbase, difficulty, gaslimit, number, timestamp, msg, data, gas, sender, sig, value, now, tx, gasprice, origin},
	keywordstyle=\color{Plum}\bfseries,
 	classoffset=3,
 	morekeywords={assert,require},
	keywordstyle=\color{red}\bfseries,
}

\lstdefinelanguage{cvl}{
	commentstyle=\color{Gray},
	morecomment=[l]{//},
	morecomment=[s]{/*}{*/},
	classoffset=0,
        escapechar=\$,
	morekeywords={anonymous, assembly, balance, break, call, callcode, case, catch, class, constant, continue, constructor, contract, debugger, default, delegatecall, delete, do, else, emit, event, experimental, export, external, false, finally, for, function, gas, if, implements, import, in, indexed, instanceof, interface, internal, is, length, library, log0, log1, log2, log3, log4, memory, modifier, new, payable, pragma, private, protected, public, pure, push, return, returns, revert,  selfdestruct, send, solidity, storage, struct, suicide, super, switch, then, this, throw, transfer, true, try, typeof, using, value, view, while, with, addmod, ecrecover, keccak256, mulmod, ripemd160, sha256, sha3},
	keywordstyle=\color{NavyBlue}\bfseries,
	classoffset=1,
	morekeywords={address, bool, byte, bytes, bytes1, bytes2, bytes3, bytes4, bytes5, bytes6, bytes7, bytes8, bytes9, bytes10, bytes11, bytes12, bytes13, bytes14, bytes15, bytes16, bytes17, bytes18, bytes19, bytes20, bytes21, bytes22, bytes23, bytes24, bytes25, bytes26, bytes27, bytes28, bytes29, bytes30, bytes31, bytes32, enum, int, int8, int16, int24, int32, int40, int48, int56, int64, int72, int80, int88, int96, int104, int112, int120, int128, int136, int144, int152, int160, int168, int176, int184, int192, int200, int208, int216, int224, int232, int240, int248, int256, mapping, string, uint, uint8, uint16, uint24, uint32, uint40, uint48, uint56, uint64, uint72, uint80, uint88, uint96, uint104, uint112, uint120, uint128, uint136, uint144, uint152, uint160, uint168, uint176, uint184, uint192, uint200, uint208, uint216, uint224, uint232, uint240, uint248, uint256, var, void, ether, finney, szabo, wei, days, hours, minutes, seconds, weeks, years},
	keywordstyle=\color{blue},
	classoffset=2,
	morekeywords={block, blockhash, coinbase, difficulty, gaslimit, number, timestamp, msg, data, gas, sender, sig, value, now, tx, gasprice, origin},
	keywordstyle=\color{Plum}\bfseries,
 	classoffset=3,
 	morekeywords={invariant,rule,assert,satisfy,require},
	keywordstyle=\color{red}\bfseries,
}

%% file: macro.tex
\newcommand{\ifempty}[3]{%
  \ifthenelse{\isempty{#1}}{#2}{#3}%
}

\newcommand{\mypar}[1]{\vspace{-0.25cm}\paragraph*{#1}}

\def\negcaptionspace{\vspace{-10pt}}

\usepackage{refcount}

\newcommand{\myfootnotetext}[1]{\footnotetext{#1\label{fn:text}%
        \edef\fnmark{\getpagerefnumber{fn:mark}}%
        \edef\fntext{\getpagerefnumber{fn:text}}%
        \ifx\fnmark\fntext\else\ClassWarning{}{footnote mark and text on different pages!}\fi}}

\newcommand{\solcode}[1]{\lstinline[language=solidity,basicstyle=\footnotesize\ttfamily]{#1}}

\newcommand{\Eg}{E.g.\@\xspace}
\newcommand{\eg}{e.g.\@\xspace}
\newcommand{\ie}{i.e.\@\xspace}

\DeclareMathSymbol{:}{\mathord}{operators}{"3A}

\newcommand{\cvc}{cvc5\xspace}
\newcommand{\ztre}{Z3\xspace}

\newif\ifliq
\liqtrue


%% file: abstract.tex
\begin{abstract}
Smart contracts are an attractive target for attackers, as evidenced by a long history of security incidents.
A current limitation of smart contract verification tools is that they are not really effective in expressing and verifying liquidity properties regarding the exchange of crypto-assets: for example, is it true that in every reachable state a user can fire a sequence of transactions to withdraw a given amount of crypto-assets?
We propose Solvent, a tool aimed at verifying these kinds of properties, which are beyond the reach of existing verification tools for Solidity. 
We evaluate the effectiveness and performance of Solvent through a common benchmark of smart contracts.
\end{abstract}

%% file: intro.tex
\section{Introduction}


In recent years we have seen a steady rise of smart contracts that implement financial ecosystems on top of public blockchains, and control today more than 100 billions of dollars worth of crypto-assets~\cite{defillama}.
The peculiarities of the setting (\ie, the absence of intermediaries, the immutability of code after deployment, the quirks in smart contract languages) make smart contracts an appealing target for attackers, as bugs might be exploited to steal crypto-assets or just cause disruption.
This is witnessed by a long history of attacks, which overall caused losses exceeding 6 billions of dollars~\cite{Chaliasos24icse}.

Formal methods provide an ideal defense against these attacks, since they enable the creation of tools to detect bugs in smart contracts before they are deployed.
Indeed, smart contracts verification tools, often based on formal methods, have been mushrooming in the last few years: for Ethereum alone --- largely the main smart contract platform --- dozens of tools exist today~\cite{Kushwaha22access}.
%
Still, the actual effectiveness of these tools in countering real-world attacks is debatable: indeed, attacks to smart contracts have continued to proliferate, refining their strategies from exploits of known vulnerability patterns to sophisticated attacks to the contracts' business logics.
As a matter of fact, the vast majority of the losses due to real-world attacks are caused by logic errors in the contract code~\cite{Chaliasos24icse}, which are outside the scope of most vulnerability detection tools. 
In particular, several real-world attacks were based on \emph{liquidity} weaknesses of smart contracts, which were exploited by attackers to steal or freeze crypto-assets~\cite{Alois17parity}.
Liquidity (also called \emph{solvency}~\cite{Kirstein21} or \emph{enabledness}~\cite{Schiffl24fmbc})
expresses the ideal behaviour of contracts in terms of the exchange of crypto-assets~\cite{Tsankov18ccs,BMZ22lmcs,Laneve23jlap}:
users want to be guaranteed that, whenever certain states are reached, they can always perform some actions that lead to a desirable assets transfer.
There are two key points in this notion: the user wants to be able to constrain \emph{who} can perform these actions, and \emph{how many} actions are needed~\cite{Schiffl24fmbc}.

The current tools that support the verification of Solidity cannot verify general liquidity properties.
This is due both to the design choices of Solidity, their target language, and to the complex logical structure of such properties.
The difficulty of verifying Solidity is caused by several glitches of high-level abstractions over the low-level target (\eg reentrancy~\cite{ABC17post}, gas costs~\cite{Grech20cacm}, and non-native tokens~\cite{Xia22sigmetrics}).
Indeed, it has been observed that existing tools already face challenges (in terms of soundness and completeness) in the verification of even simpler properties~\cite{BFMPS24fmbc}. 
We believe that effectively verifying liquidity requires first to purify Solidity from its main semantical quirks.  
Once we have done this, we must deal with the peculiar logical structure of liquidity properties. 
Indeed, liquidity cannot be expressed in terms of safety or liveness, and bounded liveness cannot model liquidity either (see ~\Cref{sec:liqVsSafLive}).
To deal with liquidity, verification techniques that go beyond safety and (bounded) liveness are therefore necessary.


\paragraph{Contributions.}

We propose \solvent, a tool that verifies liquidity properties of smart contracts. 
\solvent takes as input a contract, written in a purified version of Solidity, and a set of user-defined liquidity
properties. 
The tool translates them into SMT constraints~\cite{SMTHandbook}, reducing the verification problem to an SMT-based symbolic model checking one~\cite{SMTHandbookMC}. Then, techniques such as bounded model checking and predicate abstraction are employed, relying on Z3~\cite{z3} and cvc5~\cite{cvc5} as back-end SMT solvers.
Experiments on a benchmark of real-world smart contracts show that \solvent can efficiently verify relevant liquidity properties of their behaviour. These properties are currently out of the scope of industrial verification tools operating on the full Solidity, like \eg SolCMC~\cite{AltS22cav} and Certora~\cite{certora}, as well as academic tools~\cite{Kalra18ndss,Hajdu19vstte,Permenev20sp,Nelaturu23tdsc,Stephens21sp,Wesley22vmcai} (we discuss such tools \iftoggle{arxiv}{in~\Cref{sec:related}}{in~\cite{solvent-arxiv}}).
Solvent provides developers with useful feedback, by detecting logical errors that would otherwise remain unnoticed.
In particular, when \solvent detects a property violation, it produces a concrete execution trace that leads the contract to a state from which the desired asset exchange is unrealisable. 
%

Summing up, the main contributions of the paper include:
\begin{itemize}
\item a fully automated encoding of an expressive subset of Solidity and of liquidity properties of smart contracts into SMT constraints;
\item a toolchain to perform bounded model checking and predicate abstraction using Z3 and cvc5 as off-the-shelf SMT solvers, producing  counterexamples (that are actually replayable in Ethereum) when the property is violated;  
\item a thorough evaluation of the tool effectiveness and performance on a benchmark of real-world smart contracts, which we extend with relevant 
liquidity properties (available on the tool \href{\solventgithub}{github repository}~\cite{solvent-github});
\item a concrete demonstration of the tool applicability, showing subtle bugs in existing smart contracts that cause crypto-assets to get frozen forever.
\end{itemize}



%% file: liqVsSafLive.tex
\section{On liquidity}
\label{sec:liqVsSafLive}


In this section, we clarify our notion of liquidity, and we support our claim that it cannot be expressed in terms of safety or (bounded) liveness. 
Recall that safety has the form ``\emph{$p$ always holds}'', liveness has the form  ``\emph{$p$ eventually holds}'', and bounded liveness 
``\emph{in at most $m$ steps $p$ holds}''. 
Liquidity, instead, has the form ``\emph{for certain users $a$, there always exists a sequence of at most $m$ actions that $a$ can perform to make $p$ hold}'' (where $m$ is a parameter).
%


To illustrate the differences among these notions, consider a user opening a bank account. The user wants to be guaranteed that, if they deposit money in the bank, then they will always be able to withdraw their money by performing a \emph{reasonable amount of actions}. 
They do not want to \emph{always} withdraw their money (safety), or to just \emph{eventually} withdraw their money (liveness). 
For example, requiring liveness but not liquidity would allow the bank to give a guarantee such as ``\emph{any client will withdraw their money after the bank has been visited overall 1 million times after the deposit}'', making the liveness property ``\emph{eventually I will withdraw}'' true (assuming the fairness condition that visits to the bank happen infinitely often). 
However, the more desirable liquidity property ``\emph{I am always able to perform at most 3 actions that make me withdraw}'' would not hold. 
Bounded liveness would not be an appropriate requirement either. 
Indeed, assume that the bank guarantees that 
``\emph{the user always receives their money before 3 actions have been made}''.
This would be undesirable, since it says that \emph{any} sequence of 3 actions, performed by \emph{any} user, would trigger the withdrawal.


In LTL, we have that safety properties have the form ``${G} p$'', and liveness properties have the form ``${GF} p$''. Liquidity properties, however, cannot be expressed in LTL, as we need to existentially quantify over the paths
(``\emph{there exists a path} where the user performs at most $m$ actions to make $p$ hold''). In CTL, the closest formulation would be ``${AG\ EX\ p \lor AG\ EX\ EX\ p \lor \dots \lor AG\ (EX)^m\ p}$''. This, however, still does not capture the strategic aspect of liquidity, which requires that the sequence that makes $p$ hold consists of actions made by certain users, and therefore we need to impose conditions over transition variables. 

%% file: tool.tex
\section{Verifying liquidity properties with \solvent}
\label{sec:tool}


We demonstrate our tool through a simple example and provide some highlights on how it works. 
Solvent operates in two steps:
\begin{inlinelist}[(1)]
\item given as input a smart contract and a set of liquidity properties, it encodes the contract and the properties into constraints in the SMT-LIB standard~\cite{BarFT-RR-17};
\item then, it issues satisfiability queries to an SMT solver 
to detect if the required properties are violated. If so, it produces a counterexample, in the form of a sequence of transactions leading to a state where a required property cannot be satisfied.
\end{inlinelist}



To illustrate Solvent, we consider in~\Cref{lst:crowdfund-bug} a simple crowdfunding contract.
The contract is akin to a class in OO programming, with attributes that define its state and methods (triggered by blockchain transactions) that update it.
The user who fires the transaction (denoted by \solcode{msg.sender}) can transfer some amount (\solcode{msg.value}) of cryptocurrency to the contract along with the call. 
The constructor specifies the owner of the crowdfunding campaign, the deadline for donations, and the target amount. 
The method \solcode{donate} allows anyone to donate any amount before the deadline;
\solcode{wdOwner} allows the owner to redeem the whole contract balance if the campaign target has been reached and the deadline has expired;   
finally, \solcode{wdDonor} allows donors to withdraw their donations after the deadline, if the campaign target has not been reached.

\begin{listing}[t]
\centering
\scalebox{1}{
\begin{lstlisting}[language=solidity,caption={A crowdfunding contract (with a subtle bug) and a liquidity property.},label={lst:crowdfund-bug}]
contract Crowdfund {
  int immutable end_donate; // last block number for donations
  int immutable target;     // threshold for successful campaign
  address immutable owner;  // beneficiary of the campaign
  mapping (address => int) donors; // records users' donations
  bool target_reached;      // initialized to false 

  constructor(address o, int e, int t) { 
    owner = o; end_donate = e; target = t 
  }
  function donate() payable {
    require (block.number<=end_donate);
    donors[msg.sender] = donors[msg.sender] + msg.value;
    if (balance>=target) { target_reached = true }
  }
  function wdOwner() {
    require (block.number>end_donate && target_reached);
    owner.trasfer(balance) // send contract balance to owner
  }
  function wdDonor() { // SUBTLE BUG HERE
    require (block.number>end_donate && balance<target); 
    msg.sender.transfer(donors[msg.sender]); 
    donors[msg.sender] = 0
  }
}
property donor_wd {
  Forall xa [ !target_reached && block.number>end_donate
    -> Exists tx [1,xa] // 1 = max tx length, xa = tx sender
    [ <tx>xa.balance    // balance of xa after tx is performed 
      >= xa.balance + donors[xa] ] ]
}
\end{lstlisting}}
 \vspace{-5pt}
\end{listing}

Solvent encodes each method into an SMT constraint that, given transaction variables, ties next-state variables to current-state variables, using auxiliary variables to represent intermediate internal states. Each \solcode{require} is encoded as an if-then-else, where, if the condition fails, the method reverts, \ie next-state variables coincide with current-state variables. 

A crucial property of crowdfunding contracts is that donors can redeem their donations after the deadline whenever the target is not reached. 
We specify this property as \solcode{donor_wd} in~\Cref{lst:crowdfund-bug}.
This reads as follows: for all users \mbox{\solcode{xa},} if the target has not been reached and the deadline has passed, then there exists a sequence \solcode{tx} of transactions of length 1 signed by \solcode{xa} such that, in the state reached after executing \solcode{tx}, the balance of \solcode{xa} is increased by \solcode{st.donors[xa]}.
Solvent detects that this property is violated. This is correct, although  surprising, because of a subtle bug in \solcode{wdDonor}.
There, the \solcode{require} ensures that donors can withdraw only if the contract balance is less than the target.
This would seem correct, since \solcode{donate} is the only method that can receive ETH (as stated by the \solcode{payable} tag).
The quirk is that contracts can receive ETH even when there are no \solcode{payable} methods, through block rewards, which can send ETH to any address, or \emph{selfdestruct}, which transfer the remaining ETH in a contract to an address at their choice~\cite{solcmc-contract-balance}.
Notably, an attacker could exploit a \emph{selfdestruct} to freeze all the funds in the contract, preventing donors from withdrawing! 

To translate \solcode{donor_wd} into an SMT constraint, Solvent considers its negation, and introduces a new existentially quantified variable for \solcode{xa}, and new universally quantified variables for all transaction variables in the sequence \solcode{tx} and for all next-state variables.
Then, it reduces to checking whether there exists an \solcode{xa} for which, if the antecedent holds, for all transaction variables and next-state variables, either the transactions invalidly tie current and next-state, or the required consequent does not hold. If this formula is \emph{un}satisfiable, then the property holds; otherwise, a counterexample is given. 
\Eg, for \solcode{donors_wd}, the counterexample is the following sequence of transitions:
\begin{lstlisting}[language=solidity]
[1] constructor(2,0,2)  msg.sender=address(4)   msg.value=0
[2] donate()            msg.sender=address(4)   msg.value=1
[3] selfdestruct()      msg.sender=address(0)   msg.value=1
\end{lstlisting}
Here, the last transition represents a call to an adversarial contract, with a method invoking \emph{selfdestruct} on \solcode{Crowdfund}.%
\footnote{Note that \emph{selfdestruct} is still active~\cite{selfdestruct}.} 
This can be easily translated into a concrete Proof-of-Concept (PoC), leading the contract to a state where \solcode{donors_wd} cannot be satisfied.

To fix the contract, we replace the condition \solcode{balance<target} in \solcode{wdDonor} with \solcode{!target_reached}.
With this fix, Solvent correctly detects that \solcode{donor_wd} holds. 

Note that the fixed contract still has a liquidity vulnerability: if the target is not reached, donors can redeem their donations, but any extra funds  possibly existing at contract creation, or received through block rewards or \emph{selfdestruct} actions, will be frozen in the contract. 
To fix this vulnerability, we first need to quantify these extra funds: we do this by adding a variable \solcode{tot_donations} that we update in \solcode{donate} and in \solcode{wdDonor}, so that the extra budget is now given by \solcode{balance-tot_donations}.
We then add a method that allows anyone to transfer any extra budget to the owner after the deadline.     
Solvent verifies that the revised contract still enjoys \solcode{donor_wd}, and transfers any extra budget to the owner, \ie:
\begin{lstlisting}[language=solidity]
property no_frozen_funds {
  Forall xa [ balance>tot_donations && block.number>end_donate
  -> Exists tx [1, xa]
  [ (<tx>balance[owner] >= balance[owner] 
                           + (balance - tot_donations)) ] ]   }
\end{lstlisting}

%% file: evaluation.tex
\section{Evaluation}
\label{sec:evaluation}

\newcommand{\xmark}{\ding{55}}%

\newcommand{\statuslive}{\checkmark}
\newcommand{\statusnotlive}{\xmark}

\newcommand{\live}[1]{\ifempty{#1}{\statuslive}{\statuslive(#1)}}
\newcommand{\notlive}[1]{\statusnotlive(#1)}
\newcommand{\timeout}{T/O}
\newcommand{\tliveupto}{---}


We test our tool over a common benchmark for Solidity verification~\cite{observant}, which includes a representative set of real-world contracts and properties.
Since this benchmark is focussed on current verification tools for Solidity, which do not deal with general liquidity properties, we extend it with relevant properties of this kind for each contract (see the \href{\solventgithub}{github page}).
Overall, we end up with \vertasks verification tasks, which we manually check for the ground truth.

\mypar{Setup.}

We run \solvent on each verification task on a 3GHz 64-bit Intel Xeon Gold 6136 CPU and a GNU/Linux OS (x86\_64-linux) with 64 GB of RAM, with either \cvc (v.~1.1.3-dev.152.701cd63ef) or \ztre (v.~4.13.0) as a back-end. The run-time limit for each veriﬁcation task is 400s of CPU time.
A subset of the results are shown in Table \ref{tab:experiments} (see \href{\experimentsgithub}{github} for the full results). We mark each property as: ``\notlive{N}'', if the solver finds a trace that violates the property (with $N$ being the length of the shortest trace leading to a violation); ``\live{}'', if it proves that the property holds in all possible states; ``\live{N}'', if it proves that the property holds for every trace of length at most $N$; and ``?'' if it timeouts.


\mypar{Results.} 
First, we note that both solvers never return an inconsistent answer. 
For all non-liquid properties, except two, at least one of the solvers is able to find a counterexample. 
When a counterexample is found, the result is returned quite quickly, and the  trace is quite short. 
For liquid properties, the solvers are able to prove the property only for some instances.
Still, in most cases, they manage to verify the property up-to traces of significant length. 
Two contracts (``Payment splitter'' and ``Vesting wallet'') are significantly tough for both solvers. This is not surprising though, as they both present non-linear behaviour, thus requiring to solve SMT formulas in the theory of Nonlinear Integer Arithmetic,  which is undecidable and notoriously hard to deal with in practice for SMT solvers.\footnote{Strategies to overcome the obstacles posed by NIA have been recently discussed, for the specific case of formulas coming from the verification of smart contracts, in \cite{HozzovaNIA}.}

\mypar{Discussion.} 

The results show that our tool is particularly good
at finding counterexamples.
They are witnessed by a sequence of transactions that can be replayed in the actual Ethereum, leading to a state from which the desired outcome is unreachable. 
On the other hand, when Solvent states that a property holds, there is no guarantee that the property is preserved ``as-is''. For instance, reentrancy vulnerabilities (which are abstracted away in our symbolic semantics), can falsify the property. Nonetheless, the output of Solvent guarantees that no conceptual error has been made in the business logic of the contract.
Even when Solvent outputs $\live{N}$, the larger the $N$ the more relevant the information given to users: indeed, empirically we observed that in the benchmark~\cite{observant}, property violations are already observable after short traces.
%
%
%
Remarkably, we spot that several contracts in the benchmark~\cite{observant} have liquidity vulnerabilities, \ie, crypto-assets remain frozen in the contract (in~\Cref{tab:experiments}, where \solcode{no_frozen_funds} is \statusnotlive).  


\begin{table}[t!]
\small
\begin{center}
\scalebox{0.8}{
\begin{tabular}{|c|c|c|c|c|c|c|}
\hline
\multirow{2}{*}{\bf Contract} & \multirow{2}{*}{\bf Property} & \multirow{2}{*}{\bf Liquid?} & \multicolumn{2}{c|}{\bf \cvc} & \multicolumn{2}{c|}{\bf \ztre}  
\\
\cline{4-7}
& & & {\bf Result} & {\bf Time }   & {\bf Result } & {\bf Time }  
\\
\hline
\multirow{3}{*}{Auction} 
& \solcode{no_frozen_funds} & \statusnotlive &  \notlive{3} & 1.10 & \notlive{3} & 1.15 \\
& \solcode{seller_wd}  & \statuslive & \  \live{10} & \tliveupto &\live{9} & \tliveupto  \\
& \solcode{old_winner_wd} & \statuslive &  \live{21} & \tliveupto & \live{9} & \tliveupto \\
\hline

\multirow{2}{*}{Bank} 
& \solcode{deposit_not_revert} & \statuslive & \live{} & 2.04 & \live{} & 19.79 \\
& \solcode{withdraw_not_revert} & \statuslive & \live{7} & \tliveupto & \live{8} & \tliveupto \\
\hline

\multirow{4}{*}{Bet} 
& \solcode{any_timeout_join} & \statuslive & \live{17} & \tliveupto  & \live{12} & \tliveupto \\
& \solcode{oracle_win} & \statuslive & \live{14}  & \tliveupto & \live{7} & \tliveupto  \\
& \solcode{any_timeout_win} & \statuslive & \live{39} & \tliveupto & \live{9} &  \tliveupto \\
& \solcode{no_frozen_funds} & \statusnotlive & \notlive{2} & 0.70 &\notlive{2} & 0.74  \\
\hline

\multirow{2}{*}{Crowdfund (bug)} 
& \solcode{owner_wd}  & \statuslive & \live{} & 2.31 & \live{} & 26.64 \\
& \solcode{donor_wd}  & \statusnotlive & \notlive{3} & 1.38 & \notlive{3} & 7.78 \\
\hline

\multirow{2}{*}{\; Crowdfund (fix2) \;} 
& \solcode{owner_wd}  & \statuslive & \live{} & 2.34 & \live{} & 29.34 \\
& \solcode{donor_wd}  & \statuslive & \live{7} & \tliveupto & \live{6} & \tliveupto \\
& \solcode{no_frozen_funds}  & \statuslive & \live{9} & \tliveupto & \live{8} & \tliveupto \\
\hline


\multirow{4}{*}{Escrow} 
& \solcode{arbiter_wd_fee}  & \statuslive & \live{17} & \tliveupto & \live{10} & \tliveupto
\\
& \solcode{buyerorseller_wd_deposit} & \statuslive & \live{16} & \tliveupto & \live{41} &  \tliveupto\\
& \solcode{anyone_wd} & \statusnotlive & \notlive{2} & 0.70 & \notlive{2} & 0.76 \\
& \solcode{no_frozen_funds} & \statusnotlive & \notlive{3} & 1.17 & \notlive{3} & 1.23   \\
\hline

\multirow{3}{*}{HTLC} 
& \solcode{owner_wd}  & \statuslive & \ \live{16} & \tliveupto  & \live{9} &  \tliveupto \\
& \solcode{verifier_wd_timeout}  & \statuslive  & \live{17} & \tliveupto  & \live{9} & \tliveupto  \\
& \solcode{no_frozen_funds}  & \statuslive & \live{} & 2.37 &\live{} & 54.84 \\\hline

\multirow{4}{*}{\begin{tabular}{c} \\[-5pt] Lottery \\ \iftoggle{arxiv}{(\ref{sec:lottery})}{} \end{tabular}}
& \solcode{one_player_win} &  \statuslive  & \live{13}  & \tliveupto   &  \live{9} & \tliveupto  \\
  & \solcode{p1_redeem_nojoin} & \statuslive  & \live{23} & \tliveupto & \live{8} & \tliveupto \\
  & \solcode{p1_redeem_noreveal} & \statuslive  & \live{16} & \tliveupto  & \live{9}& \tliveupto  \\
  & \solcode{p2_redeem_noreveal} & \statuslive  & \live{17} & \tliveupto & \live{8} & \tliveupto \\
\hline

\multirow{3}{*}{Payment splitter} 
& \solcode{anyone_wd_ge}  & \statuslive & \live{2} & \tliveupto & \live{1} & \tliveupto   \\
& \solcode{anyone_wd_releasable} & \statuslive &  \live{2} & \tliveupto & \live{1} & \tliveupto   \\
& \solcode{anyone_wd} & \statusnotlive &  \notlive{1} & 0.40 &  \notlive{1} & 0.38   \\
\hline

\multirow{3}{*}{Vault} 
& \solcode{fin_owner} & \statuslive & \live{14} & \tliveupto & \live{11} & \tliveupto \\
& \solcode{canc_recovery} & \statuslive & \live{20} & \tliveupto & \live{10} & \tliveupto  \\
& \solcode{wd_fin_owner} & \statusnotlive & \notlive{1} & 0.45 & \notlive{1}& 0.44 \\
\hline

\multirow{6}{*}{Vesting wallet} 
& \solcode{owner_wd_expired}  & \statuslive & \live{} & 2.16  & ? & \timeout   \\
& \solcode{owner_wd_started}  & \statusnotlive & ? & \timeout & ? & \timeout \\
& \solcode{owner_wd_uncond}  & \statusnotlive & ? & \timeout & \notlive{1} & 0.36  \\
& \solcode{owner_wd_beforestart}  & \statusnotlive & ? & \timeout & \notlive{1} & 0.35  \\
& \solcode{owner_wd_empty}  & \statusnotlive & ? & \timeout & \notlive{1} & 0.35 \\
& \solcode{owner_wd_released}  & \statusnotlive & ? & \timeout & ? & \timeout  \\
\hline
\end{tabular}
 } 
\end{center}
\negcaptionspace
\caption{Solvent benchmark (subset; execution times are in seconds). For $\statuslive{(N)}$ properties we write ``---'' when the prover timeouts before proving $\statuslive{(N+1)}$.}
\label{tab:experiments}
\vspace{0pt}
\end{table}

%% file: related.tex
\section{Related work}
\label{sec:related}

The Ethereum ecosystem includes several bug detection tools that can spot \emph{specific} forms of liquidity vulnerabilities in smart contracts.
In that setting, these vulnerabilities are often referred to as ``Locked Ether'', which are roughly described as the absence of a mechanism to withdraw Ether from the contract.
The tools capable of detecting Locked Ether bugs include   
Slither~\cite{slither},
SmartCheck~\cite{smartcheck},
Maian~\cite{maian},
Securify2~\cite{Tsankov18ccs},
ConFuzzius~\cite{confuzzius}
and sFuzz~\cite{sfuzz}.
Each of these tools has its own internal encoding of the Locked Ether property, making it difficult to compare them~\cite{Sendner24sp}.
There are two main differences between these tools and ours.
First, while these tools can only spot that a predefined hard-coded Locked Ether property is violated, Solvent can verify custom liquidity properties, defined by our specification language.  In particular, Solvent can verify properties that express \emph{who} can withdraw \emph{how much}, and under \emph{which conditions}; not just that Ether does not get frozen in the contract.
Second, bug detection tools only focus on property violations, while Solvent (a formal verification tool) is capable of verifying that a liquidity property is satisfied (always, or up to a certain bound on the number of transactions).

Other state-of-the-art formal verification tools for Solidity do not support general liquidity properties.
SolCMC, the prover shipped with the Solidity compiler, as well as other verification tools such as Zeus~\cite{Kalra18ndss}, solc-verify~\cite{Hajdu19vstte}, VerX~\cite{Permenev20sp} and SmartACE~\cite{Wesley22vmcai}, only support safety properties, while other tools, such as Certora~\cite{certora}, VeriSolid~\cite{Nelaturu23tdsc}, and SmartPulse~\cite{Stephens21sp}, can also verify liveness properties.
However, as discussed in Section \ref{sec:liqVsSafLive}, this is not enough for liquidity properties (see \iftoggle{arxiv}{Appendix~A}{\cite{solvent-arxiv}} for a more in-depth discussion).

%


%% file: conclusions.tex
\section{Conclusions and Future work}
\label{sec:conclusions}

Solvent has already proven useful to spot general liquidity vulnerabilities, which are beyond the reach of current tools.
Still, there is space for improvements.

First, Solvent uses off-the-shelf SMT solvers, relying on bounded model checking to find counterexamples, and on predicate abstraction to prove that the property holds. 
An alternative approach that we plan to investigate is to leverage advanced techniques (\eg, abstraction-refinement, k-induction, etc.) used by modern infinite-state symbolic model checkers.
%

Second, we plan to extend our Solidity fragment to narrow the gap with the actual Solidity (see \iftoggle{arxiv}{Appendix~A}{\cite{solvent-arxiv}} for a discussion of the main differences). 
In particular, we note that concretising our \solcode{transfer} into contract-to-contract calls, as in Solidity, would require to take failures and reentrancy into account.
\Eg, consider a contract with a method:
\begin{lstlisting}[language=solidity]
foo() { owner.transfer(1); msg.sender.transfer(1); }
\end{lstlisting}
and a property requiring that, whenever the contract balance is at least 2, any user can increase their balance by 1.
In our Solidity fragment, this property is true, since both transfers will  succeed under the given hypotheses.
This is not the case in the concrete Solidity: \eg, \solcode{owner} could be a contract address, whose fallback method could fail under certain conditions. In this case, the second transfer in \solcode{foo} would not happen, and so the property would be violated.
Since considering each transfer as potentially failing would make most contracts illiquid, a possible approach would be to allow queries to specify which transfers or addresses to be considered trusted.   

Other future work include the automatic transformation of the counterexamples given by Solvent into an executable PoC (\eg, in the \href{https://github.com/NomicFoundation/hardhat}{HardHat} tool), and 
the formalisation of liquidity in a suitable logic.

%% file: ack.tex
\paragraph*{Acknowledgments.}

Work partially supported by project SERICS (PE00000014)
under the MUR National Recovery and Resilience Plan funded by the
European Union -- NextGenerationEU, and by PRIN 2022 PNRR project DeLiCE (F53D23009130001).

%% file: app-intro.tex
\section{Supplementary material}

Because of space constraints, we include in this Appendix the following supplementary material:
\begin{itemize}

\item in~\Cref{sec:solidity} we give a brief overview of Solidity, and discuss the main differences with the fragment supported by Solvent; 

\item in~\Cref{sec:solcmc-certora} we discuss related work on bug detection and formal verification tools for smart contracts. In particular, we provide a thoroughly discussion of the differences between Solvent and the two main verification tools for Solidity, \ie SolCMC and Certora, highlighting why verification of liquidity properties is currently out of their scope. We also discuss other verification tools, such as VeriSolid~\cite{Nelaturu23tdsc} and SmartPulse~\cite{Stephens21sp}, that can verify liveness properties but not liquidity ones.

\item in~\Cref{sec:lottery} we evaluate Solvent on a more complex use case, \ie a fair 2-players lottery.
Implementing this use case is quite error-prone, since the contract must implement a commit-reveal protocol that prescribes punishments whenever a player behaves dishonestly, \eg by refusing to perform some required action.
The contract must ensure that, even in these cases, an honest player has at least the same payoff that she would have by interacting with another honest player.
Proving relevant liquidity properties of this contract seems far beyond the capabilities of existing verification tools. 
To overcome these limitations, we provide Solvent with an extended syntax to express hashed timelock protocols, similarly to what is done in other smart contract languages (\eg, in \href{https://tezos.gitlab.io/alpha/timelock.html}{Tezos}).

\end{itemize}

%% file: solidity.tex
\subsection{Background on Solidity, and relation to Solvent's fragment}
\label{sec:solidity}

Solidity was one of the first contract languages to be introduced, and it is currently the main high-level smart contract language for Ethereum and other
blockchains that support the Ethereum Virtual Machine (EVM).
Solidity code is compiled into EVM bytecode, and then executed by the blockchains nodes.

Solidity follows the account-based stateful model: namely, the ownership of crypto-assets of both users and contracts is recorded into accounts; furthermore, contracts can store data (\ie, state variables) in their account. 
Accounts are partitioned into Externally Owned Accounts (EOAs), which are controlled by users through private keys, and contract accounts. Every account is uniquely identified by an \emph{address}.

Transactions are sent from EOAs to contract accounts to trigger state transitions in contracts; these transitions possibly involve transfers of crypto-currency from the caller EOA to the called contract, and from the latter to other accounts (both EOAs and contracts). 
Each transaction is signed by a single EOA, denoted as \solcode{msg.sender}, and can transfer units of ETH from the caller EOA to the contract. The amount of transferred ETH is denoted by \solcode{msg.value}.
Contracts have state variables and methods that can update them, as exemplified by the crowdfund contract in~\Cref{sec:tool}.


The main differences between the languages supported by Solvent and the actual Solidity are the following:
\begin{itemize}

\item Solidity features contract-to-contract calls, while Solvent only features EOA-to-contract calls. Contract calls in Ethereum are quite burdensome, since the callee can in turn call any other contract (including the original caller), paving the way to so-called reentrancy attacks~\cite{ABC17post}. 
While it would be possible to extend our verification technique to take contract-to-contract calls into account, in the current version of Solvent we have opted to drop this feature, since existing verification tools for Solidity already provide effective defence against reentrancy attacks.

\item In Solidity, transfers of crypto-currency from a contract to another account are encoded as contract-to-contract calls, while in Solvent we  assume that the transferred amount actually arrives at the destination address, \ie that this address is either an EOA or a contract account that does not perform a further call.
This assumption simplifies stating properties about the funds transferred from a contract (see~\Cref{sec:solcmc-certora}), and is used in other verification tools such as VerX~\cite{Permenev20sp}, where it is called 
\emph{effective external callback freedom} assumption.

\item Solidity features (possibly unbounded) loops and a complex gas mechanism (specified at the EVM level) to avoid divergent computations and reward blockchain nodes for processing transactions. Although these features are not present in Solvent, thereby limiting its expressiveness, the benchmark in~\Cref{tab:experiments} shows that Solvent is still expressive enough for a wide range of applications. Furthermore, we note that unbounded loops are discouraged even in Solidity, since they may be exploited by attackers to make a contract become stuck because an iteration exceeds the block gas limit.

\item Solvent features a special syntax to express \emph{hashed timelock protocols}, where a user first store the hash of a chosen secret in the contract, and then reveal the secret, making the contract check that the hash of the revealed secret corresponds to the stored hash.  
This is a common pattern, used \eg in gambling games and blind auctions, but where existing verification tools for Solidity fail to give the expected results (see~\Cref{sec:solcmc-certora}).
We exemplify this feature to design a 2-players lottery in~\Cref{sec:lottery}, showing that Solvent manages to provide developers with a useful feedback.  

\end{itemize}

%% file: app-related.tex
\subsection{Verification of liquidity in other tools}
\label{sec:solcmc-certora}

We now discuss more concretely why Certora, SmartPulse and VeriSolid cannot express general liquidity properties.

First, we discuss Certora, one of the leading formal verification tools for Solidity.
Consider the \solcode{Freezable} contract in~\Cref{lst:freezable-deposit}.
The contract allows anyone to withdraw part of its balance through the method \solcode{pay}, unless the variable \solcode{frozen} is true.
This variable is controlled by the \solcode{owner} through the method \solcode{freeze}: hence, the \solcode{owner} at any time can freeze the contract balance, preventing anyone from withdrawing.
A desirable property of the \solcode{Freezable} contract, and of smart contracts in general, is that crypto-assets cannot be frozen forever.

The rules in~\Cref{lst:certora:liquid-nd} are tentative specifications of the liquidity property is the Certora Verification Language (CVL).
We claim that neither rule correctly encodes the intended liquidity property:
\begin{itemize}

\item The rule \solcode{liq_satisfy} is satisfied if there exists some starting state such that,
for some \solcode{sender} address and some \solcode{v},
\solcode{sender} can fire a transaction \mbox{\solcode{pay(v)}} that increases its balance by \solcode{v}.
This is not a correct way to encode our liquidity property: indeed, Certora says that the property is satisfied, since there \emph{exists} a trace that makes the condition in the \lstinline[language=cvl]{satisfy} statement true: this is the trace where the \solcode{owner} has not set \solcode{frozen} yet. 

\item The rule \solcode{liq_assert}, which is identical to \solcode{liq_satisfy} but for the \lstinline[language=cvl]{satisfy} statement that replaces the \lstinline[language=cvl]{assert}, is satisfied if, for all reachable states, for all \solcode{sender} and \emph{for all} values \solcode{v}, a transaction \solcode{pay(v)} is never reverted.
Also this rule does not correctly specify the intended liquidity property:
Certora would correctly state that the property is false, because there are some values \lstinline{v} that make the transaction fail (\eg, when \lstinline{v} exceeds the contract balance). 

\end{itemize}

Although in this simple example we could fix the rule \solcode{liq_assert} by requiring that the transaction is not reverted for all values \lstinline{v} less than the contract balance and when \lstinline{frozen} is false, 
in general we would like to know if there \emph{exist} parameters that make the desirable property true, which is not expressible in CVL.
Actually, in \solvent we can express exactly this property, as shown in~\Cref{lst:solvent:liquid-nd}. 


\begin{listing}[t!]
\centering
\begin{lstlisting}[language=solidity,caption={A freezable deposit contract.},label={lst:freezable-deposit}]
contract Freezable {
  address immutable owner;
  bool frozen;
  
  constructor () payable {
    owner = msg.sender;
  }
  
  function freeze() external {
    require (msg.sender == owner);
    frozen = true;
  }
  
  function withdraw(int amount) external {
    require(!frozen);
    msg.sender.transfer(amount);
  }
}
\end{lstlisting}

\begin{lstlisting}[language=cvl,caption={Wrong encodings of a liquidity property in Certora.},label={lst:certora:liquid-nd}]
// Certora specification
rule liq_satisfy(address sender, uint v) {
  mathint b0 = bal(sender); // sender initial balance
  env e;
  require e.msg.sender == sender;
  withdraw(e, v);
  mathint b1 = bal(sender); // sender balance after pay(v)
  satisfy(b1 == b0 + v);    // looking for a positive example
}

rule liq_assert(address sender, uint v) {
  mathint b0 = bal(sender); // sender initial balance
  env e;
  require e.msg.sender == sender;
  withdraw(e, v);
  mathint b1 = bal(sender); // sender balance after pay(v)
  assert(b1 == b0 + v);     // looking for a negative example
}
\end{lstlisting}

\begin{lstlisting}[language=solidity,caption={Encoding of a liquidity property in \solvent.},label={lst:solvent:liquid-nd}]
// Solvent specification
property liq {
  Forall xa [
    !frozen
    -> Exists tx [1, xa]
      [ <tx>xa.balance == xa.balance + balance ]
  ]
}
\end{lstlisting}
\end{listing}

Another kind of properties that are not easily expressible are those that concerning transfers of crypto-currency from the contract. 
For example, consider the contract \solcode{Transfer} in~\Cref{lst:transfer:contract}.
The method \solcode{withdraw} allows anyone to transfer any fraction of the contract balance to the \solcode{rcv} address.
Properly formalising this simple property is surprisingly burdensome.

A first attempt would be to express the property is through the \solcode{invariant} in~\Cref{lst:transfer:solcmc}, which asserts a constraint on the balances of the contract and of \solcode{rcv} before and after a \solcode{withdraw}.
This however would not be a correct formalisation, for multiple reasons.
First, a contract that sends 10 units of crypto-currency to an intermediary who forwards the funds to the address \solcode{rcv} would violate the intended property but possibly satisfy the invariant in~\Cref{lst:transfer:solcmc}.
Second, in general the invariant might not hold, as detected by both SolCMC and Certora.
Actually, if \solcode{rcv} is a contract address, the transfer could trigger another call that in turns transfers the crypto-currency elsewhere, thus breaking the invariant%
\footnote{The actual feasibility of this further transfer also depends on the amount of gas units transferred to \solcode{rcv} and consumed by its fallback function. However, these gas costs are usually not taken into account by verifiers.}. 

A case where we are certain that the \solcode{withdraw(v)} will successfully increase the recipient's balance by \solcode{v} units of crypto-currency is when \solcode{rcv} is an EOA.
However, even in this simple case expressing and verifying the correct transfer property is problematic.
First, there is no general way for a contract to discriminate between an EOA and a contract address%
\footnote{\url{https://docs.openzeppelin.com/contracts/4.x/api/utils\#Address}}.
Second, even in the cases where it is possible to determine that an address is an EOA, existing verification tools such as SolCMC and Certora do not manage to verify that the property holds~\cite{BFMPS24fmbc}. 

In \solvent, we encode the property as in~\Cref{lst:transfer:solvent}, by specifying a constraint on the balances of the contract and of the address \solcode{rcv}. 
\solvent correctly detects that the property holds.
The underlying assumption here is that all the addresses to which a contract transfers crypto-currency behave as EOAs, \ie they cannot perform internal calls to send the received crypto-currency to some other address. Note that the \solcode{transfer} command should rule internal calls, since the amount of gas forwarded to the recipient is not sufficient to pay the gas to complete the execution of an internal call. 
The other underlying assumption is that the recipient is not rejecting inbound transfers of crypto-currency (this would be possible, \eg, by crafting a recipient contract with a fallback function that always fails). Since there is no rational reason to implement this behaviour, we assume that contracts never deliberately refuse to receive crypto-currency. Furthermore, this would be pointless, since \eg there is no way to prevent a \emph{selfdestruct} transaction to transfer crypto-currency to an address. 

\begin{listing}
\begin{lstlisting}[language=solidity,caption={A simple deposit contract.},label={lst:transfer:contract}]
contract Transfer {
  address payable immutable rcv;

  constructor() payable {
    rcv = payable(msg.sender);
  }
  
  function withdraw(uint v) public {
    require(v<=address(this).balance);
    rcv.transfer(v);
  }
}
\end{lstlisting}
\begin{lstlisting}[language=solidity,caption={A wrong attempt to reason about transfers in SolCMC.},label={lst:transfer:solcmc}]
// Invariant to be processed by the SolCMC verifier 
function invariant(uint v) public {
  uint s0 = rcv.balance;
  uint c0 = address(this).balance;

  withdraw(v);
  
  uint s1 = rcv.balance;
  uint c1 = address(this).balance;   
  
  assert(s1==s0+v && c1==c0-v);
}
\end{lstlisting}
\begin{lstlisting}[language=solidity,caption={Liveness of transfers in \solvent.},label={lst:transfer:solvent}]
// Solvent specification
property liquidity_live {
  Forall xa
  [
    true
    ->
    Exists tx [1, xa]
    [
      <tx>rcv.balance == rcv.balance + balance
    ]
  ]
}
\end{lstlisting}
\end{listing}

Besides Certora, only SmartPulse~\cite{Stephens21sp} and VeriSolid~\cite{Nelaturu23tdsc} seem capable to deal with liveness properties. 
SmartPulse targets directly Solidity code, and it has a property specification language based on Linear Temporal Logic (LTL).
This makes it possible to express liveness properties, but not liquidity properties, that are out of the reach of LTL (see~\Cref{sec:liqVsSafLive}).
Therefore, the properties supported by SmartPulse and \solvent have uncomparable expressiveness.
To be usable in practice, liveness properties must be accompanied by a \emph{fairness assumption}, \ie another LTL formula that specifies the traces where a user performs some required action in order to reach the desired state.
For example, in our crowdfunding contract (\Cref{sec:tool}) a target property could be ``if the target is not reached, then a donor gets their money back'', and the associated fairness assumption could be ``the donor performs the action \solcode{wdDonor}''.
A main difference between \solvent and SmartPulse is that, while in SmartPulse the designer of the property must anticipate, in the fairness assumption, the sequence of transactions that a user must perform in order to reach a certain state change, in \solvent this sequence is inferred by the the SMT solver. In particular, the solver infers the (minimal) length of the sequence, the called methods, and their actual parameters in order to produce the desired state change.

VeriSolid takes as input a Solidity contract and its properties expressed in Computation Tree Logic (CTL), transforms the contract into an equivalent Abstract State Machine (ASM), and verifies the properties against the ASM using tools in the BIP toolchain, such as the {nuXmv} symbolic model checker~\cite{Bliudze15atva}.
The liveness properties specified in~\cite{Nelaturu23tdsc} are not accompanied by fairness assumptions (unlike SmartPulse), but in principle this seems doable without reworking the verification techniques.%
\footnote{A recent extension of VeriSolid include some forms of fairness constraints~\cite{Chahoki23overlay}.}
We have already noted that the liveness properties expressed in CTL cannot encompass the general liquidity properties addressed by \solvent. 
We further note that to express liquidity we mix universal and existential quantification on variables (\eg, ``for all users, there exists a sequence of transactions made by the user'').

%% file: lottery.tex
\subsection{Use case: a 2-players lottery}
\label{sec:lottery}

Consider a lottery where 2 players bet 1 ETH each, and the winner --- who is chosen fairly between the two players --- redeems the whole pot.
Since smart contract are deterministic and external sources of randomness (\eg, random number oracles) might be biased,  
to achieve fairness we follow a commit-reveal-punish protocol, 
where both players first commit to the secret hash, then reveal their secrets (which must be preimages of the committed hashes), and finally the winner is computed as a fair function of the secrets.

We show below an implementation of the lottery protocol in Solvent; we then apply our tool to verify some relevant liquidity properties.   
Intuitively, the protocol followed by honest players is the following:
\begin{enumerate}
\item \solcode{player1} joins the lottery by paying 1 ETH and committing to a secret;
\item \solcode{player2} joins the lottery by paying 1 ETH and committing to another secret;
\item if \solcode{player2} has not joined, \solcode{player1} can redeem her bet after block \solcode{end_commit}; 
\item once both secrets have been committed, \solcode{player1} reveals the first secret;
\item if \solcode{player1} has not revealed, \solcode{player2} can redeem both players' bets after block \solcode{end_reveal}; 
\item once \solcode{player1} has revealed, \solcode{player2} reveals the secret;
\item if \solcode{player2} has not revealed, \solcode{player1} can redeem both players' bets after block \solcode{end_reveal+100};
\item once both secrets have been revealed, the winner, who is determined as a function of the two revealed secrets, can redeem the whole pot of 2 ETH.
\end{enumerate}

\begin{figure}[t!]
\begin{lstlisting}[language=solidity,caption={A lottery contract (part 1).},label={lst:lottery1}]
contract Lottery {
  address player1
  address player2
  int immutable end_commit	// last round to join
  int immutable end_reveal	// last round to reveal
  hash hashlock1
  hash hashlock2
  secret secret1
  secret secret2
  int state

  constructor(int tc, int tr) {
    require (tc < tr);
    end_commit = tc;
    end_reveal = tr;
    state = 0 // next = join1
  }
  function join1(address a1, hash h1) payable {
    require (state==0 && msg.value==1);
    player1 = a1;
    hashlock1 = h1;
    state = 1 // next = join2 or redeem1_nojoin
  }
  function join2(address a2, hash h2) payable {
    require (state==1 && msg.value==1);
    player2 = a2;
    hashlock2 = h2;
    state = 2 // next = reveal1
  }
  function reveal1(secret s1) {
    require (state==2 && block.number >= end_commit);
    require (sha256(s1) == hashlock1);
    secret1 = s1;
    state = 4 // next = reveal2 or redeem2_noreveal
  }
  function reveal2(secret s2) {
    require (state==4 && block.number >= end_reveal + 100);
    require (sha256(s2) == hashlock2);
    secret2 = s2;
    state = 5 // next = win
  }
  function win() {
    require (state==5);
    if ((length(secret1) + length(secret2)) % 2 == 0) {
        player1.transfer(balance)
    } else {
        player2.transfer(balance)
    };
    state = 3 // next = end
  }
  ...
\end{lstlisting}
\end{figure}

\begin{figure}[t!]
\begin{lstlisting}[language=solidity,caption={A lottery contract (part 2).},label={lst:lottery2}]
contract Lottery {
  ...
  function redeem1_nojoin() {
    require (state==1 && block.number>=end_commit);
    player1.transfer(balance);
    state = 3 // next = end
  }
  function redeem2_noreveal() {
    require (state==2 && block.number >= end_reveal);
    player2.transfer(balance);
    state = 3 // next = end
  }
  function redeem1_noreveal() {
    require (state==4 && block.number >= end_reveal+100);
    player1.transfer(balance);
    state = 3 // next = end
  }
  function empty() {
    require (state == 3);
    msg.sender.transfer(balance)
  }
}
\end{lstlisting}
\end{figure}

We implement the lottery protocol in~\Cref{lst:lottery1,lst:lottery2}.
The expected liquidity properties of the contract, formalised in~\Cref{lst:lottery-properties}, are the following:
\begin{itemize}

\item \solcode{p1_redeem_nojoin}: in state $1$, \solcode{player1} can redeem at least her bet after the block \solcode{end_commit};

\item \solcode{p2_redeem_noreveal}: in state $2$, \solcode{player2} can redeem at least both players' bets after the block \solcode{end_reveal};

\item \solcode{anyone_liquid3}: in state $3$, anyone can withdraw the whole contract balance;

\item \solcode{p1_redeem_noreveal}: in state $4$, \solcode{player1} can redeem at least both players' bets after the block \solcode{end_reveal};

\item \solcode{one_player_win}: in state $5$, either \solcode{player1} or \solcode{player2} can redeem at least both players' bets.

\end{itemize}

\noindent
Solvent correctly manages to verify that all these properties hold (up-to a given bound of transactions).
\bartnote{qui per esprimere bene la proprietà farebbe comodo avere che le transazioni in tx possano essere eseguite da player1 o da player2. Con questa estensione, si potrebbe chiedere che dallo stato 2 player1 e player2 possono fare una sequenza tx che svuota il contratto. Si può fare di meglio?}

\begin{figure}[t!]
\begin{lstlisting}[language=solidity,caption={Ideal properties of the lottery contract.},label={lst:lottery-properties}]
property player1_can_redeem_nojoin {
  Forall xa [
    st.state == 1 && st.block.number >= st.end_commit
    -> Exists tx [1, xa] [
      (<tx>player1.balance >= player1.balance + 1)  
    ]
  ]
} 
property player2_can_redeem_noreveal {
  Forall xa [
    st.state == 2 && st.block.number >= st.end_reveal
    -> Exists tx [1, xa] [
      (<tx>player2.balance >= player2.balance + 2)  
    ]
  ]
}
property anyone_liquid3_live {
  Forall xa [
    st.state == 3
    -> Exists tx [1, xa] [
      (<tx>xa.balance >= xa.balance + balance)  
    ]
  ]
}
property player1_can_redeem_noreveal {
  Forall xa [
    st.state == 4 && st.block.number >= st.end_reveal+100
    -> Exists tx [1, xa] [
      (<tx>player1.balance >= player1.balance + 2)  
    ]
  ]
} 
property one_player_win {
  Forall xa [
    state == 5
    -> Exists tx [1, xa] [
      (<tx>player1.balance >= player1.balance + 2) || 
      (<tx>player2.balance >= player2.balance + 2))
    ]
  ]
}
\end{lstlisting}
\end{figure}

%% file: main.bbl
\begin{thebibliography}{10}
\providecommand{\url}[1]{\texttt{#1}}
\providecommand{\urlprefix}{URL }
\providecommand{\doi}[1]{https://doi.org/#1}

\bibitem{solcmc-contract-balance}
{SMTChecker} and formal verification: contract balance.
  \url{https://docs.soliditylang.org/en/v0.8.24/smtchecker.html#contract-balance}
  (2023)

\bibitem{observant}
An open benchmark for evaluating smart contracts verification tools.
  \url{https://github.com/fsainas/contracts-verification-benchmark} (2024)

\bibitem{Alois17parity}
Alois, J.: Ethereum {Parity} hack may impact {ETH} 500,000 or \$146 million.
  \url{https://www.crowdfundinsider.com/2017/11/
  124200-ethereum-parity-hack-may-impact-eth-500000-146-million/} (2017),
  accessed on April 9, 2024

\bibitem{AltS22cav}
Alt, L., Blicha, M., Hyv{\"{a}}rinen, A.E.J., Sharygina, N.: {SolCMC}:
  {Solidity} compiler's model checker. In: Computer Aided Verification. LNCS,
  vol. 13371, pp. 325--338. Springer (2022). \doi{10.1007/978-3-031-13185-1_16}

\bibitem{ABC17post}
Atzei, N., Bartoletti, M., Cimoli, T.: A survey of attacks on {Ethereum} smart
  contracts {(SoK)}. In: Principles of Security and Trust ({POST}). LNCS, vol.
  10204, pp. 164--186. Springer (2017). \doi{10.1007/978-3-662-54455-6_8},
  \url{http://dx.doi.org/10.1007/978-3-662-54455-6_8}

\bibitem{cvc5}
Barbosa, H., Barrett, C.W., Brain, M., Kremer, G., Lachnitt, H., Mann, M.,
  Mohamed, A., Mohamed, M., Niemetz, A., N{\"o}tzli, A., Ozdemir, A., Preiner,
  M., Reynolds, A., Sheng, Y., Tinelli, C., Zohar, Y.: cvc5: {A} versatile and
  industrial-strength {SMT} solver. In: Int. Conf. on Tools and Algorithms for
  the Construction and Analysis of Systems ({TACAS}). LNCS, vol. 13243, pp.
  415--442. Springer (2022). \doi{10.1007/978-3-030-99524-9_24}

\bibitem{BarFT-RR-17}
Barrett, C., Fontaine, P., Tinelli, C.: {The SMT-LIB Standard: Version 2.6}.
  Tech. rep., Department of Computer Science, The University of Iowa (2017),
  available at \url{https://smtlib.cs.uiowa.edu/language.shtml}

\bibitem{SMTHandbookMC}
Barrett, C., Tinelli, C.: Satisfiability Modulo Theories, pp. 305--343.
  Springer International Publishing, Cham (2018).
  \doi{10.1007/978-3-319-10575-8_11},
  \url{https://doi.org/10.1007/978-3-319-10575-8_11}

\bibitem{SMTHandbook}
Barrett, C.W., Sebastiani, R., Seshia, S.A., Tinelli, C.: Satisfiability modulo
  theories. In: Handbook of Satisfiability (2021).
  \doi{10.3233/978-1-58603-929-5-825}

\bibitem{solvent-github}
Bartoletti, M., Ferrando, A., Lipparini, E., Malvone, V.: {Solvent: liquidity
  verification of smart contracts},
  \url{https://github.com/AngeloFerrando/Solvent}

\bibitem{BFMPS24fmbc}
Bartoletti, M., Fioravanti, F., Matricardi, G., Pettinau, R., Sainas, F.:
  Towards benchmarking of {Solidity} verification tools. In: Formal Methods in
  Blockchain ({FMBC}). OASIcs, vol.~118, pp. 6:1--6:15. Schloss Dagstuhl -
  Leibniz-Zentrum f{\"{u}}r Informatik (2024). \doi{10.4230/OASICS.FMBC.2024.6}

\bibitem{BMZ22lmcs}
Bartoletti, M., Lande, S., Murgia, M., Zunino, R.: Verifying liquidity of
  recursive {Bitcoin} contracts. Log. Methods Comput. Sci.  \textbf{18}(1)
  (2022). \doi{10.46298/LMCS-18(1:22)2022}

\bibitem{Bliudze15atva}
Bliudze, S., Cimatti, A., Jaber, M., Mover, S., Roveri, M., Saab, W., Wang, Q.:
  Formal verification of infinite-state {BIP} models. In: Automated Technology
  for Verification and Analysis ({ATVA}). LNCS, vol.~9364, pp. 326--343.
  Springer (2015). \doi{10.1007/978-3-319-24953-7\_25}

\bibitem{Chahoki23overlay}
Chahoki, A.Z., Roveri, M., Amyot, D., Mylopoulos, J.: Revisiting formal
  verification in {VeriSolid}: An analysis and enhancements. In: Workshop on
  Artificial Intelligence and Formal Verification, Logic, Automata, and
  Synthesis. {CEUR} Workshop Proceedings, vol.~3629, pp. 55--60. CEUR-WS.org
  (2023)

\bibitem{Chaliasos24icse}
Chaliasos, S., Charalambous, M.A., Zhou, L., Galanopoulou, R., Gervais, A.,
  Mitropoulos, D., Livshits, B.: Smart contract and {DeFi} security: Insights
  from tool evaluations and practitioner surveys. In: {IEEE/ACM} International
  Conference on Software Engineering ({ICSE}). pp. 60:1--60:13. {ACM} (2024).
  \doi{10.1145/3597503.3623302}, \url{https://doi.org/10.1145/3597503.3623302}

\bibitem{z3}
De~Moura, L., Bj\o{}rner, N.: Z3: an efficient {SMT} solver. In: Int. Conf. on
  Tools and Algorithms for the Construction and Analysis of Systems ({TACAS}).
  LNCS, vol.~4963, pp. 337--340. Springer (2008).
  \doi{10.1007/978-3-540-78800-3\_24}

\bibitem{defillama}
Defillama. \url{https://defillama.com/}, accessed on April 9, 2024

\bibitem{selfdestruct}
{EIP-4758}: Deactivate {SELFDESTRUCT}.
  \url{https://eips.ethereum.org/EIPS/eip-4758}, accessed on June 15, 2024

\bibitem{slither}
Feist, J., Grieco, G., Groce, A.: Slither: a static analysis framework for
  smart contracts. In: Workshop on Emerging Trends in Software Engineering for
  Blockchain, ({WETSEB@ICSE}). pp. 8--15 (2019).
  \doi{10.1109/WETSEB.2019.00008}

\bibitem{Grech20cacm}
Grech, N., Kong, M., Jurisevic, A., Brent, L., Scholz, B., Smaragdakis, Y.:
  {MadMax}: analyzing the out-of-gas world of smart contracts. Commun. {ACM}
  \textbf{63}(10),  87--95 (2020). \doi{10.1145/3416262}

\bibitem{Hajdu19vstte}
Hajdu, {\'{A}}., Jovanovic, D.: solc-verify: {A} modular verifier for
  {Solidity} smart contracts. In: Verified Software. Theories, Tools, and
  Experiments ({VSTTE}). LNCS, vol. 12031, pp. 161--179. Springer (2019).
  \doi{10.1007/978-3-030-41600-3\_11}

\bibitem{HozzovaNIA}
Hozzov{\'a}, P., Bend{\'i}k, J., Nutz, A., Rodeh, Y.: Overapproximation of
  non-linear integer arithmetic for smart contract verification. In:
  International Conference on Logic for Programming, Artificial Intelligence
  and Reasoning. EPiC Series in Computing, vol.~94, pp. 257--269 (2023).
  \doi{10.29007/h4p7}

\bibitem{certora}
Jackson, D., Nandi, C., Sagiv, M.: Certora technology white paper.
  \url{https://docs.certora.com/en/latest/docs/whitepaper/index.html} (2022)

\bibitem{Kalra18ndss}
Kalra, S., Goel, S., Dhawan, M., Sharma, S.: {ZEUS:} analyzing safety of smart
  contracts. In: Network and Distributed System Security Symposium ({NDSS}).
  The Internet Society (2018)

\bibitem{Kirstein21}
Kirstein, U.: Formal verification helps finding insolvency bugs — {Balancer
  V2} bug report.
  \url{https://medium.com/certora/formal-verification-helps-finding-insolvency-bugs-balancer-v2-bug-report-1f53ee7dd4d0},
  accessed on August 30, 2024

\bibitem{Kushwaha22access}
Kushwaha, S.S., Joshi, S., Singh, D., Kaur, M., Lee, H.N.: Ethereum smart
  contract analysis tools: A systematic review. IEEE Access  \textbf{10},
  57037--57062 (2022). \doi{10.1109/ACCESS.2022.3169902}

\bibitem{Laneve23jlap}
Laneve, C.: Liquidity analysis in resource-aware programming. J. Log. Algebraic
  Methods Program.  \textbf{135},  100889 (2023).
  \doi{10.1016/J.JLAMP.2023.100889}

\bibitem{Nelaturu23tdsc}
Nelaturu, K., Mavridou, A., Stachtiari, E., Veneris, A.G., Laszka, A.:
  Correct-by-design interacting smart contracts and a systematic approach for
  verifying {ERC20} and {ERC721} contracts with {VeriSolid}. {IEEE} Trans.
  Dependable Secur. Comput.  \textbf{20}(4),  3110--3127 (2023).
  \doi{10.1109/TDSC.2022.3200840}

\bibitem{sfuzz}
Nguyen, T.D., Pham, L.H., Sun, J., Lin, Y., Minh, Q.T.: {sFuzz}: an efficient
  adaptive fuzzer for {Solidity} smart contracts. In: International Conference
  on Software Engineering ({ICSE}). pp. 778--788. {ACM} (2020).
  \doi{10.1145/3377811.3380334}

\bibitem{maian}
Nikolic, I., Kolluri, A., Sergey, I., Saxena, P., Hobor, A.: Finding the
  greedy, prodigal, and suicidal contracts at scale. In: Annual Computer
  Security Applications Conference ({ACSAC}). pp. 653--663. {ACM} (2018).
  \doi{10.1145/3274694.3274743}

\bibitem{Permenev20sp}
Permenev, A., Dimitrov, D.K., Tsankov, P., Drachsler{-}Cohen, D., Vechev, M.T.:
  {VerX}: Safety verification of smart contracts. In: {IEEE} Symposium on
  Security and Privacy. pp. 1661--1677. {IEEE} (2020).
  \doi{10.1109/SP40000.2020.00024}

\bibitem{Schiffl24fmbc}
Schiffl, J., Beckert, B.: A practical notion of liveness in smart contract
  applications. In: Formal Methods in Blockchain ({FMBC}). OASIcs, vol.~118,
  pp. 8:1--8:13. Schloss Dagstuhl - Leibniz-Zentrum f{\"{u}}r Informatik
  (2024). \doi{10.4230/OASICS.FMBC.2024.8}

\bibitem{Sendner24sp}
Sendner, C., Petzi, L., Stang, J., Dmitrienko, A.: Large-scale study of
  vulnerability scanners for {Ethereum} smart contracts. In: {IEEE} European
  Symposium on Security and Privacy ({S{\&}P}). pp. 220--220. {IEEE} (2024).
  \doi{10.1109/SP54263.2024.00230}

\bibitem{Stephens21sp}
Stephens, J., Ferles, K., Mariano, B., Lahiri, S.K., Dillig, I.: {SmartPulse}:
  Automated checking of temporal properties in smart contracts. In: {IEEE}
  Symposium on Security and Privacy. pp. 555--571. {IEEE} (2021).
  \doi{10.1109/SP40001.2021.00085}

\bibitem{smartcheck}
Tikhomirov, S., Voskresenskaya, E., Ivanitskiy, I., Takhaviev, R., Marchenko,
  E., Alexandrov, Y.: {SmartCheck}: Static analysis of {Ethereum} smart
  contracts. In: Workshop on Emerging Trends in Software Engineering for
  Blockchain ({WETSEB}). pp. 9--16. {ACM} (2018). \doi{10.1145/3194113.3194115}

\bibitem{confuzzius}
Torres, C.F., Iannillo, A.K., Gervais, A., State, R.: {ConFuzzius}: {A} data
  dependency-aware hybrid fuzzer for smart contracts. In: {IEEE} European
  Symposium on Security and Privacy ({EuroS{\&}P}). pp. 103--119. {IEEE}
  (2021). \doi{10.1109/EUROSP51992.2021.00018}

\bibitem{Tsankov18ccs}
Tsankov, P., Dan, A.M., Drachsler{-}Cohen, D., Gervais, A., B{\"{u}}nzli, F.,
  Vechev, M.T.: Securify: Practical security analysis of smart contracts. In:
  {ACM} {SIGSAC} Conference on Computer and Communications Security ({CCS}).
  pp. 67--82. {ACM} (2018). \doi{10.1145/3243734.3243780}

\bibitem{Wesley22vmcai}
Wesley, S., Christakis, M., Navas, J.A., Trefler, R.J., W{\"{u}}stholz, V.,
  Gurfinkel, A.: Verifying {Solidity} smart contracts via communication
  abstraction in {SmartACE}. In: Verification, Model Checking, and Abstract
  Interpretation ({VMCAI}). LNCS, vol. 13182, pp. 425--449. Springer (2022).
  \doi{10.1007/978-3-030-94583-1\_21}

\bibitem{Xia22sigmetrics}
Xia, P., Wang, H., Gao, B., Su, W., Yu, Z., Luo, X., Zhang, C., Xiao, X., Xu,
  G.: Trade or trick?: Detecting and characterizing scam tokens on {Uniswap}
  decentralized exchange. In: {ACM} {SIGMETRICS/IFIP} {Performance}. pp.
  23--24. {ACM} (2022). \doi{10.1145/3489048.3522636}

\end{thebibliography}
